\documentstyle[12pt]{article}
\newcommand{\beq}{\begin{equation}}
\newcommand{\eeq}{\end{equation}}
\newcommand{\beeq}{\begin{eqnarray}}
\newcommand{\eeeq}{\end{eqnarray}}

\begin{document}
\begin{titlepage}
\title{\bf Fermion Condensation and
Non Fermi Liquid Behavior in a Model with
 Long Range Forces.}
\author {{\bf J. Dukelsky}$^1$,
{\bf V.A. Khodel}$^2$,
{\bf P. Schuck}$^3$, and
{\bf V.R. Shaginyan}$^4$\\
{\small
 $^1$ Instituto de Estructura de la Materia, Serrano 123,
  28006 Madrid, Spain}\\
{\small
$ ^2$ Kurchatov Institute of Atomic Energy Moscow, 123182,
 Russian Federation}\\
{\small $^3$ Institut des Sciences Nucleaires, IN2P3-CNRS, UJFG
53 Avenue des Martyrs,
F-38026 Grenoble-Cedex, France}\\
{\small $^4$ St. Petersburg  Nuclear Physics Institute,
Gatchina, 188350,
Russian Federation}}
\date{}

 \maketitle

\end{titlepage}

{\bf Abstract.}
The phenomenon of the so called Fermion condensation, a phase transition
analogous to Bose condensation but for Fermions,
postulated in the past to occur
in systems with strong momentum dependent forces,
 is reanalysed in a model
with infinite range interactions. The strongly non
 Fermi Liquid behavior of this
 system
is  demonstrated  analytically
at $T=0$ and at $T\neq 0$ in the superconducting
 and normal
phases. The validity of the
 quasiparticle picture is investigated and seems to
 hold true
for temperatures less than
the characteristic temperature $T_f$ of the Fermion
condensation.

\vskip .7cm
\section {Introduction}

Recently, proceeding from the Landau approach [1] which treats the ground
state energy $E$ of a Fermi system as a functional of its quasiparticle
distribution $n_{{\bf p},\sigma}$, Khodel and Shaginyan [2] proposed
the idea of the so called Fermion Condensation (FC).
This phenomenon can occur in systems with a sufficiently strong  momentum
dependence of the effective interaction between particles
giving rise to a downswing of the quasiparticle
dispersion in the vicinity
of the chemical potential $\mu$ (Fig.1).
 A distinct feature of this phenomenon
 is the emergence of a flat (dispersionless) portion of
extension $\Omega$ in the quasiparticle spectrum $\varepsilon_{{\bf p},
\sigma}={\delta E\over \delta n_{{\bf p},\sigma}}$,
located at the chemical potential $\mu$:
$\varepsilon_
{{\bf p},\sigma}=\mu,$ for $ {\bf p}\ \in \Omega$.
This plateau is obtained from the Maxwell
 construction as shown in Fig.1 and as
 explained in [2].
As a consequence of the high degeneracy of this spectrum the distribution
 $n_{{\bf p},\sigma}$varies with $p$
 continuously with a  finite derivative $\partial n_{{\bf p},\sigma}/
\partial  p$ in contrast to $n_F(p)=\theta(p-p_F)$
 of ordinary Fermi liquids.
A macroscopic part of quasiparticles can then
occupy the region $\Omega$ ( the Fermion condensate), a feature
very similar to Bose-Einstein condensation of bosons. The condensate
wave function turns out to be degenerate.
This degeneracy may be removed by the
interactions excluded from the initial functional $E[n_{{\bf
 p},\sigma}]$, e.g.
 pairing correlations [2]. As a result, the plateau in the spectrum
of single-particle excitations is distorted, but, as a rule, the spectrum
remains quite flat since the strength  of  pairing force is rather
small.

 Some difficulties with the
concept of FC  have later been discussed by P.Nozieres within
the Hartree-Fock approach by using a schematic model of infinite
 range forces [3] providing an extremely strong momentum dependence
of the effective interaction entering the Fock term. In fact,
 the radius $r_s$ of the force  is supposed to greatly exceed the
interparticle distance $r_0$ and
then the model [3] provides us with the leading
terms of the $r_0/r_s$ expansion.  This model proves to be very
 useful, since   it enables us to analytically evaluate most of the
basic quantities. We will adopt it to a large
extent in this paper.

The purpose of the present work is to further
elucidate  the rather unusual
properties of Fermi systems which undergo FC.
The outline of the paper is as follows. In Sect. 2 we will address  the
question of the interplay between
superconductivity and  FC at $T=0$.
Sect.3 is devoted to
finite temperatures, and in Sect.4 we discuss the validity of the
quasiparticle picture. Finally in Sect.5 we conclude our paper.

\section{\bf Interplay of superconductivity and
Fermion condensation.}

In order to formulate our studies in a convenient way we take up the
model of infinite range  forces [3] in a slightly more general way.
We start out from the following Hamiltonian
\beq
H=H_0 + H_{int},
\eeq
where
\beq
H_0=\Sigma_{p,\sigma}\xi^0_{{\bf p}}
a^+_{{\bf p},\sigma}a_{{\bf p},\sigma},
\quad \xi^0_{{\bf p}}=\varepsilon^0_{{\bf  p}}-\mu,
\eeq
and
\beq
H_{int}={ 1\over 2}\Sigma_{{\bf p}_1,{\bf p}_2,{\bf q},
 \alpha_1,..,\alpha_4}[U_0({\bf q})
\delta_{\alpha_1,\alpha_3}\delta_{\alpha_2,\alpha_4}+
U_s({\bf q})\sigma^i_
{\alpha_1,\alpha_3}\sigma^i_{\alpha_2,\alpha_4}]a^+_{{\bf p}_1,\alpha_1}
a^+_{{\bf p}_2,\alpha_2} a_{{\bf p}_1+{\bf q},\alpha_3}
a_{{\bf p}_2-{\bf q},\alpha_4}
\eeq
with
$\varepsilon^0_{{\bf p}}=p^2/2m$,  $U_0({\bf q})=U_0\delta({\bf q })$
and $U_s({ \bf q})=U_s\delta({\bf q })$. The
 constants $U_0$ and $U_s$ are  assumed to be  small  compared to the
Fermi energy $\varepsilon ^0_F=p^2_F/2m$.
The only difference with the model [3] is that the
Hamiltonian (1-3)
contains an additional  spin exchange mixture.

In order to
study the superconducting properties of our model let us calculate
the ground state energy from the usual singlet BCS wave function
\beq
|BCS>=\Pi\biggl(u_p+v_p
a^+_{{\bf p},\sigma}a^+_{-{\bf p},-\sigma}\biggr)|vac>.
\eeq
This ground state is known to be the vacuum to the quasiparticle
operators $\alpha_{{\bf p},\sigma}$ (i.e.
 $\alpha_{{\bf p},\sigma}|BCS>=0)$ which are related to the original
operators through
 the Bogoliubov transformation
\beq
  a_{{\bf p},+}=u_ p\alpha_{{\bf p},+}+v_p\alpha^+_{-{\bf p},-},\qquad
  a_{{\bf p},-}=u_p\alpha_{{\bf p},-}-v_ p\alpha^+_{-{\bf p},+},
\eeq
with the coefficients $u_p$ and $v_p$ obeying the normalization condition
$u^2_p+v^2_p=1$.
The expectation value of $H$ is easily calculated and we obtain
\beq
 F=<BCS|H|BCS>\equiv E-\mu N=\Sigma_p [2\xi^0_p n_p+
 V_1 n^2_p-V_2\kappa^*_p\kappa_p],
\eeq
where
\beq
 n_{{\bf p}}={1\over 2}
 \Sigma_{\sigma} <a^+_{{\bf p},\sigma}a_{{\bf p},\sigma}>=v^2_p
  ,\qquad \kappa_{{\bf p}}=
< a_{{\bf p},+}a_{{-\bf p},-}>=u_pv_p,
\eeq
are the occupation numbers and the pair distribution function,
respectively.
A straightforward
calculation yields
$V_1=-(U_0+3U_s)$, $V_2=3U_s-U_0$.
For the model [3] with $U_s=0$, the
condition $V_1>0$, necessary for the existence of the Fermion condensate,
  is fulfilled only  for attractive forces $U_0<0$ that also leads
  inevitably  to Cooper pairing, since in this case $V_2>0$.
 However, by adding a spin-dependent term $U_s<0$ we can, for example,
take repulsive $U_0>0$ and choose
the spin-dependent constant $U_s<0$ such
 that FC exists
 while Cooper pairing not. Note, that the Hartree term reducing to a
renormalization of the  chemical potential $\mu$ is omitted in (6).
We also want to indicate that the infinite range of $V_1$ and $V_2$
only is a
convenient tool to trigger FC and to qualitatively
investigate the interplay between pairing and FC.

 At $T=0$, the BCS extremum equation
  ${\delta F\over\delta v_p}= 0$,
 corresponding to the minimum of $E$
provided
 $V_1+V_2>0$,
 has the well known form(see, e.g. [4]):
\beq
%\varepsilon_p-\mu -\Delta_p(u^2_p-v^2_p)/ 2u_pv_p=0,
\xi_p -\Delta_p(u^2_p-v^2_p)/ 2u_pv_p=0,
\eeq
where $\xi_p=\varepsilon_p-\mu$ and
  the Landau quasiparticle energy $\varepsilon_p
= {\delta E\over \delta n_p}$
and the gap $\Delta_p
 =-{\delta F\over\delta\kappa_p}$
are given by
\beq
\varepsilon_p=\varepsilon^0_p +V_1 n_p,
\eeq
and
\beq
 \Delta_p=V_2u_pv_p.
\eeq
It is worth noting that the substitution of the Fermi step $n_p=
\theta(p-p_F)$ into eq.(9) yields a downswing
in $\varepsilon_p$ with  a vertical slope
lying exactly at the Fermi surface.

Upon substitution of (10) into (8) with the replacement of
 $u^2_p-v^2_p$ by $1-2n_p$  we obtain
\beq
 \xi_p-V_2(1-2n_p)/2=0.
\eeq
Together with (9) this yields
\beq
n_p={\mu+V_2/2-p^2/2m\over V_1+V_2},\qquad
\xi_p={p^2/2m-\mu+V_1/2\over V_1+V_2}V_2.
\eeq
From (12) we see, as discussed
in the  introduction, that $\varepsilon_p-\mu$ is
 dispersionless
if $V_2$ is zero, i.e. for vanishing pairing interaction.

The occupation numbers $n_p$ must lie between 0 and 1. We therefore have
  $n_p=1$ for $p\leq p_i$ and $n_p=0$ for $p\geq p_f$. Upon inserting these
conditions into (12) one finds \beq p_f^2=2m(\mu+V_2/2),\qquad
p_i^2=2m(\mu-V_2/2-V_1).
\eeq
With the help of (13)  $n_p$ and $\kappa_p$ can be
rewritten in a very compact form
\beq
 n_p={p^2_f-p^2\over p_f^2-p_i^2},\quad \kappa_p
 ={\sqrt{(p^2_f-p^2)(p^2-p^2_i)}
\over p_f^2-p_i^2},\quad  p_i< p < p_f,
\eeq
while outside the above domain $n_p$ coincides
with the usual Fermi gas distribution
 (see Fig.2) where $\kappa_p$ is zero.
We infer that the minimum of the
free energy $F$ is attained inside the functional
 space $[n_p]$ only if the momentum $p$ belongs to
the condensate region, otherwise
this minimum gets into a boundary point of the
 space $[n_p]$
at which the gap $\Delta_p$ vanishes.

An additional relation, linking $p_i$ to $p_f$ and appropriate for
the evaluation of the chemical potential $\mu$,
follows from the equality between the quasiparticle and particle numbers.
We have
$${p^3_F\over 3\pi^2}={p^3_i\over
3\pi^2}+\int_{p_i}^{p_f}{dp p^2\over\pi^2}
{p^2_f-p^2\over p^2_f-p^2_i}.$$
The result of the integration is
\beq
 p^3_F={2(p^5_f-p^5_i)\over 5(p^2_f-p^2_i)}.
\eeq
In this article we shall concentrate on the case
 $V_2<<V_1$ in which the pairing forces are weaker
 than the particle-hole
 interaction $V_1$ since the case $V_1\simeq V_2$ has already been
analyzed in [3] while the choice $V_1<<V_2$ leads to ordinary BCS theory.
The  quasiparticle  spectrum $E_p$ can be evaluated by
rewriting (11) as
$2n_p=1-2\xi_p/V_2$ and comparing to the standard
 BCS form
$$2n_p=1-\xi_p/E_p \eqno (12^\prime),$$
 where $E_p$ is
the quasiparticle energy. Then  we find that
the spectrum $E_p$ in the FC region
 is dispersionless as in [3]:
 \beq
 E_p=\sqrt{\xi_p^2+\Delta_p^2}
={V_2\over 2} ,\qquad  p_i<p<p_f.
 \eeq
Relation (16) can also be verified directly using (7,10,12,14). With
(16) it is  straighforward to show that the usual expression for
the pair distribution
 $$\kappa_p=
 \frac{\Delta_p}{2E_p}\eqno (14')$$
is compatible with (14).

In the case $V_2=0$,  the quasiparticle
distribution $n_p$  (12) transforms to
\beq
n_p={\mu -p^2/2m\over V_1},\quad p^0_i<p< p^0_f,
\eeq
where $p^0_f=\sqrt{2m\mu}$ and $p^0_i=\sqrt{2m(\mu -V_1)}$.In this case
the expression (17)  coincides
with that determined by the minimum equation  [2]: ${\delta E\over \delta
 n_p}=\mu$ and
the spectrum $\varepsilon_p$ from (12)
does have the plateau $\varepsilon_p=\mu$
inside the interval $p_i^0<p<p_f^0$.

Using (13), (15) and keeping
only leading order terms in
 $V_i/\varepsilon^0_F$  we get  for the correction to
the chemical potential:
 \beq
 \mu-\mu_F\simeq
-(V_1+V_2)^2/48\mu_F).
 \eeq
Here we have used the formula
$\mu_F
=p^2_F/ 2m+V_1/2$
for the HF chemical potential. With the chemical potential (18) one can
evaluate the spectrum of the collective excitations
 that are sound vibrations.
 The sound velocity $c_s$ is found from the relation
$mc^2_s=\rho {\partial \mu\over \partial\rho}$ and is given by
\beq
c^2_s={p^2_F\over 3m^2}\biggl(1 + {(V_1+V_2)^2\over
48(\mu_F)^2}\biggr).
\eeq
 We see that the rearrangement of the
single particle degrees of freedom influences little the collective ones
since the second term is a small correction to the first one.

From the  above considerations we can draw our first conclusions.
For example, let us consider the limit
 $V_2\rightarrow 0$. From ($12,12^\prime,16$)
 we see that the distribution $n_p$
 is an absolutely smooth
and continuous function of $V_2$ down to  $V_2=0$. The fall off width
of $n_p$ is practically independent of $V_2$ and is dominated by the
particle-hole component $V_1$.
 Furthermore, the superconductivity order parameter (14,14')
is also a continuous function of $V_2$ down
to the limit $V_2=0$ which thus
differs from zero even  if $\Delta_p=0$. In
other words, we can say that even in the absence of pairing forces, if
the situation for the Fermion condensation prevails, the system
spontaneously undergoes a transition to the state with
broken  gauge invariance that is traditionally associated with
superconductivity. This is also clear from the fact that, according
to what we just have said, the BCS state (4) still holds true at $V_2=0$.
(However, for $V_2=0$, the critical temperature $T_c
=0$, see Sect.3).
Another way to see this stems
from  the usual Thouless criterium for the onset of superconductivity
which we here consider using the Bethe Salpeter equation
for the 2-particle Green function
(in a symbolic writing) [4]
\beq
G_2 = G^0 + G^0V_2G_2,
\eeq
where $G^0=(1-2n_p)/( \omega -2(\varepsilon_p-\mu))$. For attractive
$V_2\neq 0$ eq.(20) yields the usual Cooper pole singularity. However,
in systems with FC, eq.(20) shows a  pole at $\omega=0$
 for the condensate state with $\varepsilon_p=\mu$
 even if $V_2\equiv 0$.

A consequence of these considerations  concerns
 the entropy $S$ of  systems with  FC. Since
the ground state wave function (4) does not change its BCS structure as a
function of $V_2$ down to $V_2=0$ and since $|BCS>$
is a pure state, we conclude
that the entropy $S$ of the system is zero even at $V_2\equiv 0$. Thus,
the entropy paradox inferred from a nonzero value of $S(T=0)$, when
 calculated  on the basis of the ordinary   Fermi gas formula
\beq
S=-2\int{d^3p\over (2\pi)^3} [n_p\ln n_p +(1-n_p)\ln (1-n_p)],
\eeq
with the distribution (12), is removed. Instead we
should use the BCS expression for  $S$
(see below) which again shows that $S=0$ at $T=0$.

The superfluid properties following from the formulas obtained, turn out
to be quite  unusual. First,
in the general case of finite range
forces when the strength of the pairing term
$V_2<<V_1$,
we seemingly deal with
the weak coupling limit of BCS theory. However, in
this case, as first  shown in [2] and also seen from
 (16), the gap in the   spectrum of
the single particle excitations has no
 BCS exponential smallness:
  $\Delta_{BCS}\sim \exp(-\varepsilon^0_F/V_2)$,
 since on the contrary, it turns out to be linear
in the particle-particle (p-p) effective coupling constant $V_2$.
Furthermore, outside the interval
$p_i<p<p_f $,
the gap $\Delta_p$ vanishes,
the boundaries $p_i$ and $p_f$
practically depending only on the value of $V_1$.
 Thus, the gap $\Delta_p$
exists only in the region occupied by the Fermion condensate.
 These properties and the fact
  that $\kappa_p\neq 0$ even for $V_2=0$ (and thus
$\Delta_p=0$) show that the strong BCS pairing is triggered by $V_1$ (in
 conjunction with
the quasiparticle plateau).
Pairing with such unusual properties
is a ``shadow" of the Fermion condensation.

A quite unexpected point is that a nontrivial solution $\Delta_p\neq 0$
exists in our frame of singlet pairing even if $V_2$ becomes negative,
i.e. repulsive, contradicting completely the Cooper pairing pattern. This
can be seen in the following way. First of all, nothing dramatic happens
if the sign of $V_2$ in $\Delta_p$ (10) and $\kappa_p (14,14^\prime)$ is
turned around. Of course, the form $V_2\kappa^2_p$ in the energy (6)
indicates a loss of energy for finite $\kappa_p$ rather than a gain
as this was  the case for $V_2>0$, i.e. attraction. However, the other
option, which is no pairing, means unambiguously that $n_p$ is back to
the Fermi step $n_F(p)=\theta(p-p_F)$. Because of the square of the
momentum distribution $n_p$ this  in turn entails an energy
loss in the term $V_1n^2_p$ even greater than if we had kept $\kappa_p$
finite ($V_2$ repulsive!). We convinced ourselves from numerical studies
that the inclusion
of the kinetic energy  does not turn over the
situation of the energy balance,
 at least for the case $|V_2|<<V_1$ considered
here. That a
gain in energy prevails even if $V_2$ is repulsive can also  be seen
from eq.(18): the sign of the correction to the chemical potential
$-(V_1+V_2)^2/48\varepsilon^0_F$ is independent of the sign of $V_2$,
as long as $|V_2|<<V_1$ ( otherwise higher order terms must be
included). Another way to characterize the situation is that FC is
analogous to Bogoliubov pairing in a nonideal Bose gas.
  Indeed, according to (10), $V_2$ is nothing
but the factor of proportionality in the relation between the gap
$\Delta_p$ and the anomal density
 $\kappa_p$. In contrast to BCS theory, $\kappa_p$
  is practically independent of the p-p interaction $V_2<<V_1$
   (see eq.(14),(14')). It is ``prepared"
    by the interaction $V_1$   entailing
 a non-Fermi-step distribution $n_p$ (see Fig.2).
\section{Finite temperature}

The extension of our model to finite temperature is, in principle,
straightforward. It follows along the lines of ordinary BCS theory [4].
First, the  quasiparticle distribution $n_p$ and the anomal density
$\kappa_p$ are modified as follows
\beq
n_p=v^2_p(1-f_p)+u^2_pf_p,\qquad \kappa_p=u_pv_p(1-2f_p),
\eeq
where
$f_p=<\alpha^+_p\alpha_p>$ is the distribution
function for the quasiparticle excitations (see below). They are described as
 a gas and the entropy $S$ is given by the integral
\beq
S=-2\int {d^3p\over (2\pi)^3}[f_p\ln f_p + (1-f_p)\ln(1-f_p)].
\eeq
We have now two different equations of the minimum:
${\delta \Omega\over \delta
v_p}=0$ and ${\delta \Omega\over \delta f_p}=0$ with $\Omega=E-\mu N-TS$.
The last one  is written as
\beq
\xi_p(1-2v^2_p)+2u_pv_p\Delta_p-T\ln{1-f_p\over f_p}=0.
\eeq
where $(\varepsilon_p-\mu)=\xi_p$ as before.
When deriving this equation
we again have used the definitions $\varepsilon_p
={\delta E\over \delta n_p}$, and $\Delta_p=-
{\delta F\over \delta \kappa_p}$.
 The first equation of the minimum keeps the same form as
at $T=0$ (eq.(8)).
 Therefore introducing the quasiparticle energy $E_p$ by a finite
temperature generalization of (16)
\beq
E_p(T)=\sqrt{\xi_p^2(T) +\Delta_p^2(T)},
\eeq
we can, as before, solve  eq.(24) yielding the   ordinary relations
for the  $u$- and $v$-factors:
\beq
v^2_p={1\over 2}\biggl(1-{\xi_p\over E_p}\biggr),\qquad
u^2_p={1\over 2}\biggl(1+{\xi_p\over E_p}\biggr).
\eeq
 Then inserting these relations together with (25)
 into (24) we are led to a seemingly ordinary Fermi-Dirac distribution
\beq
f_p={1\over e^{{E_p\over T}}+1}.
\eeq
As a matter of fact, this relation represents an  equation for $f_p$
since $E_p$ implicitly depends   on $f_p$.
\vskip 1cm
{\bf a. The case ${\bf T\leq T_c}$}.

Let us turn to the solution of the equation  for the gap $\Delta_p(T)$
following from its definition and the relation (22) for $\kappa_p(T)$
\beq
\Delta_p(T)=V_2u_p(T)v_p(T)(1-2f_p(T))
\eeq
Upon substituting into this equation relations (25-27)
we get
\beq
\Delta_p=V_2 \frac{\Delta_p}{2E_p(T)} \tanh\frac{E_p(T)}{2T}
\eeq
We shall  see that a nontrivial solution of this equation written as
\beq
2E_p(T)=V_2\tanh{E_p(T)\over 2T}, \qquad p_1(T)<p<p_2(T)
\eeq
only exists in an interval
$[p_1,p_2]$, narrower than the interval $[p_i,p_f]$ (13) of FC.
  Being independent of $p$, $E_p(T)$  drops with $T$ and
vanishes at $T=T_c=V_2/4$ [3]. This is also seen in Fig.3 where
 a graphical solution of (30) is given.

From (30) one finds $\tanh {E_p\over 2T}=2E_p/V_2$
which allows us to rewrite
(27) as
\beq f_p(T)={1\over 2}-{E_p(T)\over
V_2},\qquad p_1<p<p_2.
\eeq
 With this relation, eq.(22) for $n_p$ takes
the same form as the first of eqs.(12)
 with the only replacement of $\mu(T=0)$ by
$\mu(T)$.  Taking into account that the second relation between these
quantities following from the definition $\varepsilon_p={\delta E\over
\delta n_p} $
 stays unchanged (see (9)) we have for $n_p$ and $\varepsilon_p$
practically the same form as at $T=0$:
\beeq
n_p(T)&=&{\mu(T)+V_2/2-p^2/2m\over
V_1+V_2},\nonumber\\
 \varepsilon_p(T)&-&\mu(T)={p^2/2m-\mu(T)+V_1/2\over
V_1+V_2}V_2, \quad p_1<p<p_2.
 \eeeq

 Outside this interval the gap  $\Delta_p=0$ and the
quasiparticle distribution becomes
\beq
n_p={1\over e^{{\varepsilon_p(T)-\mu(T)\over T}}+1},\qquad
p<p_1 \quad {\rm or} \quad p>p_2,
\eeq
 as can be seen from the first of eqs.(22). Since relation (9), linking
the quasiparticle energy $\varepsilon_p$ to the distribution
 $n_p$, also holds at $T\neq 0$, this means that eq. (33)
is a quite intricate  equation  rather than an explicit
 formula for the calculation of $n_p$. We can rewrite
relation (33) between these quantities in the form
\beq
\varepsilon_p-\mu=T\ln {1-n_p\over n_p}
\eeq
 and substitute it into (9). Then one obtains
\beq
p^2/2m-\mu+V_1n_p=\varepsilon_p-\mu=
T\ln {1-n_p\over n_p}, \quad  p<p_1,\quad
 {\rm or} \quad p>p_2,
\eeq
 This is a closed equation
for the evaluation of $n_p$. Finding $n_p$ we can insert it into
eq.(34) to calculate  $\varepsilon_p$. This procedure
 will be considered in more detail for the case $T>T_c$ later on.

 Now we turn to  the calculation of
the boundaries $p_1(T)$, $p_2(T)$, determining the volume occupied by
the superfluid phase. They  are found
 from matching  the distribution $n_p$ given by the
formulas (32) and (33). At these boundaries,
eqs.(32) and (33),
 we can rewrite the matching equation in the form
\beq
-{\xi(p_k)\over V_2}+{1\over 2} =
 {1\over e^{{\xi(p_k)\over T}}+1}\quad k=1,2
\eeq
or equivalently
\beq
2\xi(p_k)= V_2\tanh {\xi(p_k)\over 2T}
\eeq
where $p_k=p_1 $ or $p_2$. It is the same equation as (30) but for
$\Delta_p =0$, as it must be.
It also has a nontrivial
solution only if $T<T_c=V_2/4$.
The distance between the points $p_1(T)$ and
$p_2(T)$ determining the volume occupied by the superfluid phase
 diminishes with $T$ and they go to meet each other at   $T=T_c$ at
  the midpoint $p_M$ of the
 interval $[p_i,p_f]$  where $n_{p_M}=1/2$.

 Upon expanding the r.h.s. of (37) which vanishes
at $T=T_c$ one finds for the support of the superfluid phase close to
$T_c$ :
\beq
p_2-p_1=4m{(V_1+V_2) T_c\over p_M V_2}
\sqrt{3\biggl(1-{T\over T_c}\biggr)}.
\eeq
The same procedure applied to eq.(30) yields
\beq
E_p=2T_c\sqrt{3\biggl(1-{T\over T_c}\biggr)},\qquad p_1<p<p_2.
\eeq
Thus, the gap in the spectrum of the single particle excitations has a
quite unusual behavior.  It is limited by the boundaries $p_1$ and $p_2$
and the support shrinks as $T$ approaches $T_c$.
 Since the absolute value
of the gap  also diminishes similar to the usual BCS gap, it is like ``an
ice cube which is melting". This behavior is shown in Fig.4. It can
easily be
obtained from $\Delta_p=\sqrt{E^2_p(T)-\xi_p^2(T)}$ using the solution of
eq.(30) shown in Fig.3.
 It is in strong contrast with the $T$-dependence
of the usual BCS gap which
 ``melts" only in ``height" but whose $p$-extension, given by the Debey
 temperature, is independent of $T$.  The  shrinkage of the volume of the
superfluid phase near $T_c$ significantly  influences  the density of
states $\rho(\varepsilon)$ and may exhibit itself in tunneling phenomena.
\vskip 1cm
{\bf b. The case ${\bf T>T_c}$}.
Let us now turn to the  normal state of the system, i.e. to $T> T_c$.
Since then $\Delta_p=\kappa_p=0$, eqs. (33,34 ) and
(35) hold in the whole domain.  It is straightforward to solve
eq.(35) with (33)  for  $\varepsilon_p$.  The result  is shown in Fig.5
for $T<<\varepsilon^0_F$  ($T_c$ can
 always be chosen arbitrarily small in the case
 $V_2<<V_1<<\varepsilon^0_F$  considered here).
We see that in comparison with
 the zero temperature case of Fig.1 the plateau is
just a little tilted and rounded off at the end points.
 Inserting $\varepsilon_p(T)$ of Fig.5 (see also eq. (41) below)
into (33) we realise that even
 for $T>T_c$ the distribution function has qualitatively
the same behavior as  the one seen in Fig.2
with only corners rounded by the
 temperature effects.
The spread of $n_p$ for $T\geq T_c$ is
therefore still governed by features of the
 Fermion
condensation and not by temperature.
Thus at $T_c<T<V_1$ we are faced with a
 quite
 interesting situation.
 Superfluid
 current is absent while phenomena which occur due to the spread of the
momentum distribution $n_p$, usually associated with superconductivity
(e.g. deviations from the Korringa
 law in the spin-lattice relaxation rate [5]), persist.

To have more
 quantitative and analytical
insight into what is going on
 let us first find the low temperature expansion
 for $n_p$ in the limit
$V_2\rightarrow 0$. For small
 $T$ the r.h.s. of eq.(35) is small. The lowest
 order of $n_p$ is
obtained in omitting the r.h.s. completely.
The result, of course, coincides
 with (17).
The next correction $n_p^{(1)}$ is found by
the substitution of (17) into the
 r.h.s. of (35) that
yields
\beq
n^{(1)}_p={\mu -p^2/2m\over V_1}+
{T\over V_1}\ln{V_1-\mu +p^2/2m\over \mu  -p^2/2m }.
\eeq
 This correction  contains a small parameter
  $T/V_1$. The iterations can be continued
 giving rise to  a low temperature expansion of
the quasiparticle distribution $n_p$ in the case $\Delta_p=0$.
There is  a one to one
correspondence between iterating
the solutions for the distribution $n_p(T)$
and the spectrum $\xi_p(T)$.
To linear order the latter  is given by (see eq.(34))
\beq
\xi_p= T\ln {1- n_p^{(0)}\over
n_p^{(0)}}=T\ln {V_1-\mu +p^2/2m\over \mu -p^2/2m
 }+O(T^2).
\eeq
In the limit $\varepsilon_p\rightarrow \mu$,
at $T$ finite, we can further
 simplify (41),
since then the argument of the logarithm must
 be close to one. This leads to
\beq
\xi_p(T)\simeq 2T\biggl[{2\xi^0_p \over V_1} +1\biggr]+O(T^2).
\eeq
This result can also be obtained more
directly upon inserting (33) into (35) and
 taking the limit
$\xi_p\rightarrow 0$ at finite $T<<V_1$. However (42)
is valid only in the
immediate vicinity of  $\varepsilon_p $
to $\mu$ whereas (41) holds for a longer
 range of $p$-values. From (41) and (42) we see that
  the quasiparticle spectrum $\varepsilon_p(T)$ in
 the region of the Fermion condensate depends linearly on $T$.
 This feature is very different from the ordinary Fermi liquid
with its $T^2$ dependence of $\varepsilon_p(T)$ stemming from
the Landau formula for the variation of the quasiparticle
energy
$$\delta \varepsilon_p(T)=2\int
 {d^3p_1\over (2\pi)^3}f({\bf p},{\bf p}_1)\delta
 n_{p_1},$$
where $f$ is the Landau scattering amplitude.
A linear temperature law may seem  quite surprising,
 since in the case considered
  the specific heat $C_V=\partial E/\partial T=T\partial S/\partial T
=-T\Sigma_{\bf p}(\varepsilon_p-\mu)\partial n_p/\partial T$
   is a linear function
 of $T$ [3] and hence
  $E(T)-E(0)\sim T^2$ as in ordinary Fermi liquids (this is easily
understandable, since for FC we have $\xi_p/T=const$ (see eq.(42)) and
thus $\partial n_p/\partial T=0$. For this reason the condensate does not
contribute to $C_V$).

Let us therefore  pin down the point where the
conventional argument leading to the $T^2$-dependence of
$\varepsilon_p(T)$ at low $T$ fails.
To this purpose, we
 differentiate both  sides of the previous equality with
respect to $T$ and obtain
\beq
{\partial \varepsilon_p(T)\over \partial T}=-2\int {d^3p_1\over (2\pi)^3}
f_0(p,p_1){\xi_{p_1}\over T^2}n_{p_1}(1-n_{p_1}).
\eeq

Here $f_0$ is the s-wave part of $f$ and use has been made of (33) for
  $dn_p/dT=(\varepsilon_p-\mu)n_p(1-n_p)/T^2$ neglecting, as usual
for normal Fermal liquids,
the insignificant $T$-dependence of the quasiparticle energies on $T$.
 One can then introduce $x_p=\xi_p/T$ as a new integration variable and
 extend the integration limit at low T with good
 accuracy to $\pm\infty$.
Since $xn(x)(1-n(x))$ is an odd function of $\xi_p$ we have to expand the
remainder of the integrand, i.e.
$\sqrt{\varepsilon_p}/(d\varepsilon_p/dp)$ to linear order in $\xi_p$
and hence:
$\partial \varepsilon_p(T)/\partial T\sim T$ or $\varepsilon_p \sim T^2$.
In a system with FC this demonstration, however, does
not hold, since $(d\varepsilon_p/dp)^{-1}$ diverges exactly at the Fermi
surface. Therefore all arguments in favor of the $T^2$ behaviour  of the
spectrum $\varepsilon_p(T)$ fail and formula (41) survives.

A straightforward comparison  demonstrates  a
remarkable correspondence between the present results and those of
the Landau theory:
most relations of ordinary Fermi liquid theory also hold  for
systems with  FC, we simply have to remember that
the effective mass diverges as $T^{-1}$:
\beq
 m^*={1\over p} \biggl({\partial\varepsilon_p\over \partial
 p}\biggr)^{-1}_{p_F}={(p^2_F-p^2_i)
(p^2_f-p^2_F)\over 2T(p^2_f-p^2_i)}\simeq m{V_1\over 4T}.
\eeq
 We will use this relation in Sect.4. (For quantities which contain the
derivative
$\partial n_p/\partial T$ like the specific heat $C_V\sim p_Fm^*T$, the
insertion of (44) is not valid, since otherwise $C_V$ would be a constant
in $T$ in contrast to what we have seen above).

The memory of  FC for temperatures beyond $T_c$
 can  be seen in a particularly transparent way from the density of
states
\beq
\rho(\varepsilon)=
2\int{d^3p\over (2\pi)^3}\delta(\varepsilon-\varepsilon_p(T)).
\eeq
This integral can be calculated analytically since $\partial
 \varepsilon_p/\partial p$
 can be obtained by differentiating eq.(35). The result is
\begin{eqnarray}
\rho(\varepsilon)&=&{V_1\over 2\pi^2}(2m)^{3/2}[\varepsilon-V_1
 n(\varepsilon-\mu)]^{1/2}[1+{V_1\over T}
n(\varepsilon-\mu)n(\mu-\varepsilon)]\nonumber\\
&\simeq &  N^o_F {p^2_f-p^2_i\over 2mT}
n(\varepsilon-\mu)n(\mu -\varepsilon),
\quad  \mu \simeq\varepsilon,
\end{eqnarray}
where $n(\xi)=(e^{\xi/T} +1)^{-1}$ and $N^0_F=p_Fm/\pi^2$.
 For  typical situations $T_c<T<V_1$ we show $\rho(\varepsilon)$
in Fig.5.

We see that the condensate contribution is spread
near the Fermi surface over
 the  interval $\delta\varepsilon\sim T$ and it is enhanced by the factor
$(p^2_f-p^2_i)/2mT\sim V_1/T>1$ if $T\sim T_c$. Its magnitude drops as
 $T^{-1}$. It becomes of the order of
the regular  contribution $N^0_F$ to the density
of states at
 a characteristic temperature,  denoted further by $T_f$ which is
\beq
 T_f\sim
{p^2_f-p^2_i\over 2m}\sim \varepsilon^0_F{\Omega_c\over \Omega_F},
\eeq
where $\Omega_c$ stands for the condensate volume and $\Omega_F$, for
the volume of the Fermi sphere. Above $T_f$
the effects of FC become insignificant.
With (47) we can write the enhancement factor
$\rho(\varepsilon=\mu)/N^o_F$
of the density of states near the Fermi surface
 in terms of $T_f$ as
\beq
{\rho(\varepsilon=\mu)\over N^o_F}\simeq{T_f\over T}.
\eeq
It is worth noting that in such a form the
estimates (47,48) are still valid  for an arbitrary form
of the effective particle-hole interaction as long as it gives rise to
 FC.  We also realise that there is not
 a definite $T$ value at which the effect of  FC
 stops, its influence diminishes gradually.

As a further peculiar point characterizing FC
we want to consider the entropy at $T>T_c$. Using (23)
in the low temperature    limit we obtain

\beq
S(T)\simeq-2\int {d^3p\over (2\pi)^3}n_p(T=0)\ln n_p(T=0)\sim
 V_1\ln{\varepsilon^0_F\over V_1}.
\eeq
 Thus, the entropy of the system
 with the Fermion condensate at $T\geq T_c$
 drastically exceeds
ordinary superconductor values $S_{BCS}(T_c)\sim T_c\sim V_2$. This is a
 specially peculiar
  ``visit card" of  FC in homogeneous systems.

Summing up this section we may say
 that systems with the Fermion condensate
exhibit a quite unusual non Fermi liquid
temperarture behavior. The main
characteristic features are that the gap as a
 function of $T$
not only shrinks in magnitude but
 also in phase space volume and that the level
 density enhancement
and flattening of the quasiparticle
 spectra, i.e. the typical Fermion condensate
features,  persist up to temperatures
 $T\simeq T_f$ far greater than $T_c$.
 Furthermore,
the entropy for $T>T_c$ is drastically
 enhanced with respect to ordinary Fermi
 liquid values.

Let us just add a  word of the behavior
 of the various quantities as a function
 of $T$ for $T<T_c$.
It is readily realized that the presence of a finite gap barely alters
the characteristic features found for $T>T_c$
indicating once again that in this kind
 of physics FC is
the driving mechanism and not pairing.
The latter is a host,
itself strongly influenced by  FC.

\section{ Validity of the quasiparticle picture}

All the results of this
 article have been obtained on the assumtion that the
approach based on the  quasiparticle picture is valid implying that
 the width $\gamma$ of the relevant
single particle states (in our case, the condensate) does not
exceed the quasiparticle energy $\varepsilon_p$. This assumption is
fulfilled for superfluid systems  with the Fermion condensate  because of
the presence of the gap in the spectrum
of the single particle excitations.
Thus, at $T<T_c$ the quasiparticle picture  is valid.

We shall see that the width $\gamma$ remains small even if $T$
exceeds $T_c$. This implies that  the quasiparticle picture
 survives  turning to  the normal state
 of these systems. Considering this problem we can
 treat it, as it was said before, within the ordinary Fermi
 liquid approach by simply taking into account  the $T$ dependence of the
 effective mass
$m^*(T)\sim T^{-1}$ (see eq.(44)). Much work has been done within
 the ordinary Fermi liquid approach to simplify the
calculations and we shall make use of well known  results. First of
 all,
the width $\gamma$ is inversely proportional to the collision time
 for which the following formula can be derived [6-8]
\beq
 {1\over \tau_0}={(m^*)^3T^2\over 8\pi^4}<{W(\theta,\phi)
\over\cos{\theta\over 2}}>,
\eeq
 where $W(\theta,\phi)$ is  the transition probability   depending  on
the angle $\theta$ between the vectors ${\bf p}_1$ and ${\bf p}_2$
of the incoming  particles and the angle $\phi$ between the planes defined
by the vectors ${\bf p}_1,{\bf p}_2$ and
${\bf p}_1^\prime,{\bf p}_2^\prime$.
 The brackets $<...>$ denote angular averaging.

 It is useful to introduce  a dimensionless quantity $w$
defined as follows: $<W/\cos{\theta\over 2}>=2\pi w/N_F^2$ where
$N_F=p_Fm^*/\pi^2$ is the density of states at the Fermi surface.
With this result
 the formula for the collision time $\tau_0$ becomes
\beq
{1\over \tau_0}={\pi\over 4}{m^*T^2\over  p^2_F}w,
\eeq
where [6-9]
\beq
w={1\over 2\pi}\int_0^{\pi}d\phi\int^{-1}_1  d\cos\theta \biggl [
{1\over 4}{|A^s(\theta,\phi)+A^a(\theta,\phi)|^2\over \cos{\theta\over
 2}}+{1\over 2}
{|A^s(\theta,\phi)-A^a(\theta,\phi)|^2\over \cos{\theta\over 2}}\biggr],
\eeq
and $A^s$ and $A^a$ are components of the
scattering amplitude. The triplet  $A_T$ and
 singlet $A_S$ scattering
 amplitudes are related to $A^s$ and $A^a$ as follows:
 $A_T=A^s+A^a$ and
$A_S=A^s-3A^a$.
 Let us note that in a Fermi liquid, without
spontaneous breaking of symmetry, the
knowledge of  the component $A^s(\theta,\phi)$  is enough to
reconstruct on the basis  of  symmetry relations the other component
 $A^a(\theta,\phi)$.
These relations are written
 for the triplet and singlet scattering amplitudes
 separately
\begin {eqnarray}
A_T({\bf p}_1,{\bf p}_2,{\bf p}_3,{\bf p}_4)=-
A_T({\bf p}_2,{\bf p}_1,{\bf p}_3,{\bf
p}_4),\nonumber\\
A_S({\bf p}_1,{\bf p}_2,{\bf p}_3,{\bf p}_4)=
A_S({\bf p}_2,{\bf p}_1,{\bf p}_3,{\bf p}_4).
\end{eqnarray}
Substituting here $A^s$ and $A^a$  leads to
\beq
A^a(\cos\theta,  q^2)=-{1\over 3}A^s(\cos\theta,  q^2)- {2\over
 3}A^s(\cos\theta,   q^{\prime 2}).
\eeq
We have used that  $\cos\theta$
remains unaltered under the interchange of the  initial momenta while the
 transferred momentum
 ${\bf q}$ is replaced by $ {\bf q^\prime}
 ={\bf p}_3-{\bf p}_2$. One has
 [8]:
${\bf q}^2=({\bf p}_3-{\bf p}_1)^2=4p^2_F\sin^2{\theta\over
 2}\sin^2{\phi\over 2}$ and
$ {\bf q}'^{2} =2p^2_F(1-\cos\theta)-q^2$.

The variables $\cos\theta$ and ${\bf q}$
 are particularly  convenient if  $A^s$ is
  linked to the Landau scattering
 amplitude $f$
 at an arbitrary value of the momentum transferred [6,9].
 The corresponding equation has the form
\beq
A^s({\bf p}_1,{\bf p}_2, {\bf q})=f({\bf p}_1,{\bf p}_2, {\bf q})+2\int
 {d^3p\over (2\pi)^3}
f({\bf p}_1,{\bf p}, {\bf q})
{n_{{\bf p}+{\bf q}/2}-n_{{\bf p}-{\bf q}/2}\over
\varepsilon_{{\bf p}+{\bf q}/2}-\varepsilon_{{\bf p}-{\bf q}/2}-\omega}
A^s({\bf p},{\bf p}_2, {\bf q}).
\eeq

This equation is, as usual, solved by expanding both sides in
 Legendre polynomials. The familiar result for each component is:
\beq
A^s_l({\bf q})={f_l({\bf q})N_F\over 1+N_Ff_l({\bf q})L({\bf q})/(2l+1)}.
\eeq
Here the dimensionless Lindhardt function $L({\bf q})=-\chi_0/N_F$
  is introduced where $\chi_0({\bf q})$
is the  linear response function of noninteracting quasiparticles
\beq
\chi_0({\bf q})=2\int
{d^3p\over (2\pi)^3}
      {n_{{\bf p}+{\bf q}/2}-n_{{\bf p}-{\bf q}/2}\over
\varepsilon_{{\bf p}+{\bf q}/2}-\varepsilon_{{\bf p}-{\bf q}/2}-\omega},
\eeq
where the energy transfer  $\omega=\varepsilon_{{\bf
 p}_1^\prime}-\varepsilon_{{\bf p}_1}$.
This function exists in   closed form [6] and varies smoothly with
 the parameters $q/p_F$ and $\omega/qv_F$.

Since nothing is known about the
Landau parameters $f_l$ with $l\geq 2$  we keep only the l=0 and l=1
harmonics of $f(\theta)$ neglecting all others with $l\geq 2$ as is
usual in  Fermi liquid theory
 [9]. In addition, we shall consider $f_0(q)$ and $f_1(q)$ as
 smooth
functions of $q$. Hence, for $T\rightarrow 0$ we shall neglect the "1"
 in the denominator of (56),
 since the density of states
diverges for $T\rightarrow 0$. Then
 the scattering amplitude $A$ can
   fully be constructed. As a result,  one obtains:
\begin{eqnarray}
A^s(\theta,\phi)={1+3\cos\theta\over
 L({\bf q})},\nonumber\\ A^a(\theta,\phi)=
-{1+3\cos\theta\over 3L({\bf q})}
-2{1+3\cos\theta\over 3L( {\bf q}')}.
\end{eqnarray}
Thus, we infer that in the strong coupling
 limit the scattering amplitude $A(\theta,\phi)$ becomes a
 universal function independent of any parameter
 characterizing the interaction between particles.

At this point one may wonder whether screening corrections to the
Landau scattering amplitude $f$ in (55) do not play a very important
role as well (this point had also been stressed in [3]).
In fact, this is not the case. A convenient frame to discuss this point
are the selfconsistent equations for $f$ derived by Babu and Brown [10].
In that work $f$ is split into a direct and an induced part. The direct
part is a Bruckner G-matrix and the induced part $f^{ind}$
corresponds to an
exchange of an (RPA) phonon between the particle and the hole
(schematically: $f=G+f^{ind}$).
For our
purpose it is sufficient to restrict  to
 the s-wave part $f_0$ of $f$. Then
the induced part is given by (eq.(4.23) and eq.(5.20) of [10]):
\beq
-f^{ind} = 2 f_0^2 \frac{N_F L({\bf q})}{ 1 + f_0 N_F L({\bf q} )}.
\eeq
As before we can neglect the "1" in the denominator of (59) and obtain
$f^{ind} = -2 f_0$ which is a finite result. Therefore screening
corrections do not invalidate our above conclusion.

As mentioned already, the $q$-dependence of the Lindhardt function is
 insignificant
 and, in  first approximation, it can be ignored by taking $L(q)=L(0)=1$.
 Then
 $w$ is evaluated analytically and after a simple integration one obtains
\beq
w={64 \over 5}.
\eeq
With this result the width $\gamma$  is calculated in  closed form:
\beq
\gamma(T)={1\over \tau_0}={16\pi\over 5}{m^*T^2\over  p^2_F}.
\eeq
Inserting here $m^*/m=T_f/4T$ from (44) we  finally find
\beq
\gamma(T)={2\pi\over 5} {T_f\over \varepsilon^0_F} T.
\eeq
 Thus, $\gamma(T)$ is a smooth function of T,
   the width of the condensate states turns out to be rather small,
 and  no inconsistency emerges in our calculations. We therefore
can infer that the quasiparticle picture turns out to be valid in all the
temperature interval $0<T<T_f$  of our interest.

It should be mentioned that, had we treated
 $\gamma$ in the Born approximation,
the quasiparticle lifetime would go to zero with $T$ as in [3].
The reason for this difference stems from
our interaction which takes
into account  screening effects.
In different words, within  perturbation
 theory  the product of the effective interaction and the density
 of states goes
 to infinity. However, if  screening effects are properly taken
 into account  this product stays  of
 the order of one in these systems.

It is useful to discuss also the $\varepsilon$-dependence of the
imaginary part ${\rm Im}\Sigma({\bf p}.\varepsilon)$ of the mass
operator $\Sigma({\bf p},\varepsilon)$ in systems with FC.
 Usually, to obtain the $\varepsilon$-dependence of
 ${\rm Im}\Sigma$, if its $T$-dependence is already known, it is
enough to replace $T$ by $\xi=\varepsilon-\mu$. This rule implies that
in the system with  FC, the imaginary part of the
mass operator is proportional to $\xi$ in contrast to the $\xi^2$-
dependence inherent in normal Fermi systems.
 A straightforward calculation
 does confirm the linearity of ${\rm Im}\Sigma$ as a
 function of $\xi$ [2].
The difference stems from the fact that, in  systems with  FC
 one of the three quasiparticles of the decay channel
 can belong to the condensate. Since the condensate
  energy is fixed, there is
 no additonal integration related to this  particle and the phase volume
 has a value as if only two-quasiparticles were present in the decay.
This
leads to the linear behavior of ${\rm Im}\Sigma({\bf p},\varepsilon)$
in $\varepsilon-\mu$.

\section{ Discussion and Conclusion}
In this article, we have
again carefully analysed the implications and the physics of the
Fermion condensation phenomenon. This phase transition was postulated to
occur in strongly correlated Fermi systems several years back [2].
Our investigations were carried out in a slightly generalized version
of the model of infinite
range forces [3]
which enabled us to analytically demonstrate the non Fermi liquid
behavior of systems with FC. In our case we  decoupled
the  link $V_1=V_2$ considered in [3] between the parameters of the
 particle-hole and particle-particle channels and analysed the case
$V_2<<V_1<<\varepsilon^0_F$.
The first thing which our study revealed is that
the BCS ground state persists even if pairing forces $(V_2)$ are absent.
The  smearing of the quasiparticle momentum distribution $n_p$ is
therefore due to the particle-hole ($V_1$) rather than the
particle-particle ($V_2$) interaction. This implies that
 the system with  FC is simultaneously in the
paired state with phase locking independent of whether there is a pairing
force or not.
 Since the BCS ground state is a pure
state there is no residual entropy problem at $T=0$ as this was evoked
earlier [3]. Even more dramatic, we found that, due to an unusual
balance, the BCS solution persists even for repulsive $V_2$ as long as
$|V_2|<<V_1$. In these limits the BCS state therefore exists independent
of the sign of $V_2$ and the entropy  always is
 zero as $T\rightarrow 0$.
At finite $T$ there exists a critical temperature $T_c=
V_2/4$ where the gap disappears. This disappearance occurs, however,
in a non-Fermi-liquid fashion: $\Delta_p$ not only shrinks in
magnitude
but also in momentum extension (phase space) which may be a measurable
effect. Once the gap has disappeared the system still shows Fermion
condensation behavior up to temperatures $T_f\sim V_1>T_c$: the typical
flat plateau in the single particle spectrum $\varepsilon_p$ is only
slightly tilted ($\sim T$) and rounded off
 at the edges and the fall off width of $n_p$ is of the order of $V_1$
 and not of the order of $T<V_1$. This is another
example of non-Fermi-liquid behavior. The effect nicely shows up in
 a strong enhancement of the level density at the Fermi level for
$T\geq T_c<T_f$. One can also say that the standard Fermi liquid
expressions remain valid with one replacement: for
$T>T_c$ the effective mass behaves itself
 as $m^*\sim T^{-1}$.

An important issue of FC
is the validity of the quasiparticle picture. This was put into question
in [3]. However, there particle collisions have been treated in the
Born approximation whereas the diverging density of states demands a
nonperturbative treatment. Including  polarization effects yields
a quasiparticle width $\gamma$ of the
 order of $T$ and the quasiparticle
concept is preserved.

Of course, the phenomena analyzed here can occur not only in systems
with infinite range forces. Solvable Hartree-Fock-like models
 suggested in [2,3] do confirm
the conclusion that  the case of finite range forces merely
changes  quantitative details without changing   qualitative
 features
of the Fermion condensation such as the plateau in $\varepsilon_p$ etc.
In particular, the shrinkage of the volume occupied by
the superfluid phase also survives with only one modification:
 the gap $\Delta_p$
acquires an exponentially small tail instead of a sharp cut off.
 No change in the conclusions
also occurs if we go beyond the Hartree-Fock approximation. For example,
we can add to the functional (7) a term of non HF nature of the form
$\delta F\sim \Sigma_{p,p_1}n^2_p n^2_{p_1}$. With this term the problem
again can be solved nearly analytically and we can verify that all
qualitative results, obtained above, hold in this model, too.
The same is true for the
  relation (44) between the characteristic temperature
$T_f$ and the size of the region occupied by the Fermion condensate.
Further, when calculating the width $\gamma$, we did not use the HF
 approach at all so that  the quasiparticle picture persists
 beyond the HF approximation.

A separate  question  concerns  the possibility of
the existence of  Fermion condensation in real physical systems.
First of all, it is
 related to the evaluation of  the single
particle spectra in such systems. A theory of these spectra in
 normal Fermi liquid based on a functional approach has been
 constructed in [2]. It is in agreement with the RPA results obtained by
Gell-Mann-Brueckner-Galitskii [11,12].
Results of  calculation of the
single particle excitation spectra carried out for
electron systems within the jellium model demonstrate that the
 point in the dispersion $\varepsilon_p$,
  where the derivative $d\varepsilon_p/dp=0$,
 a precursor of the Fermion condensation, does emerge in these systems at
the densities at which the effective interaction between
electrons becomes strong enough [13]. In this case
 one can also calculate the width $\gamma$ for the condensate particles
 and again find  $\gamma(T) \sim T$ validating
 the quasiparticle picture.

An issue
of considerable
interest is the finding [14]
within the local density approximation (LDA) of bifurcated saddle
points in double-plane materials like
YBa$_2$Cu$_3$O$_7$, YBa$_2$Cu$_4$O$_8$, and Bi$_2$Sr$_2$CaCu$_2$O$_{8+
\delta}$. It seems that, for a certain range of doping, these saddle
points are pinned to the Fermi level and that they lie exactly
where experimentally an anomalous flattening of the single particle
dispersion in two-dimensional strongly correlated electron systems
 is observed [15].
 With the Fermi level in the region
of the bifurcated saddle point again the Maxwell construction seems
applicable [16], leading to  straight segments of  Fermi lines
instead of the two-dimensional plateaus
 discussed in this work. It should,
however, be noted that these flat Fermi  lines also lead to a diverging
density of states $\rho(\varepsilon=\mu)$ [17]. We intend to investigate
this problem in a future publication.

To summarize, we have analysed the rather unusual properties of  Fermi
 systems which undergo   a phase transition to
 Fermion condensation.
 Strong non Fermi liquid
behavior has been demonstrated for those   systems whereas the
 the quasiparticle picture seems to survive: for $T<T_c$ due to the
 gap in the single particle spectrum, for $T>T_c$, due to a screening
suppression of the matrix elements responsible for the decay process.

{\bf Acknowledgements}

We are grateful to   A.I. Liechtenstein and P. Nozieres
 for extended discussions and valuable remarks. We also acknowledge
 discussions with  A.A. Abrikosov, O.K. Andersen,
F. Borsa, J.C. Campuzano,
  J.W. Clark, A.I. Larkin, R.M. Martin, D. Pines,
L.P. Pitaevskii, J. Ranninger, J.P. Vary, G. Vignale
 and M.V. Zverev.
 This work was supported in part by the
 Russian
Foundation for Fundamental Invetsigation (Grant $N_O.$
95-0204481 and Grant
No.\ 94-02-04-665a) and by the International Science Foundation
under Grant No.\ Mh 8300.J.D. acknowledges the support of DGICYT (Spain)
under contract N.PB 92/0021-C02-01.
\newpage
\begin{center}
{\bf References}
\end{center}
\vskip .3cm
\begin{itemize}
\item[1)] Landau L.D. JETP {\bf 30}, 1058 (1956)
\item[2)] Khodel V.A., Shaginyan V.R. JETP Letters {\bf 51}, 553
(1990);      Khodel V.A., Khodel V.V.,  Shaginyan V.R.
Phys. Rep. {\bf 249}, 1(1994).
\item[3)] Nozieres P. Journ. Phys. I France {\bf 2}, 443 (1992).
 (1994)
\item[4)] Fetter A.L. and Walecka J.D. Quantum Theory of Many-Particle
Systems. McGraw Hill, 1971.

Ring P. and Schuck P. Nuclear Many Body Problem. Springer-Verlag 1980.
\item [5)] Slichter C. Nuclear magnetic resonance.
Springer-Verlag, New York,  1990.
\item [6)] Pines D. and Nozieres P. The Theory of quantum liquids, W.A.
 Benjamin, New York (1966).

Baym G. and Pethick C.J. The Physics of Liquid and Solid Helium.Part II,
ch.1.  Wiley, New York, 1978.
\item [7)] Jens
en H.H., Smith H. and Wilkins J.W.
Phys. Rev. {\bf 185}, 323 (1969).
\item [8)] Lawrence W.E. and Wilkins J.W. Phys. Rev. {\bf B7},
 2315 (1973).
\item [9)] Dy K.S. and Pethick C.J. Phys. Rev. {\bf 185}, 373 (1969).
\item [10)] Babu S., Brown G.E. Ann. of Physics (N.Y.) {\bf 78},
 1 (1973).
\item [11)] Gell-Mann M. and Brueckner K.A. Phys.Rev.{\bf 106},
364 (1957).
\item [12)] Galitskii V.M. Selected Works. Moscow, Nauka, 1983.
\item [13)] Zverev M.V., Khodel V.A. and Shaginyan V.R. JETP (in press).
\item [14)] Andersen O.K., Jepsen O., Liechtenstein A.I., Mazin I.I.
Phys. Rev. {\bf B49}, 4145 (1994).
\item [15)] Dessau D.S. and Shen Z.-X. Phys. Reps. {\bf 253}, 1 (1995).
\item [16)] Khodel V.A. Shaginyan V.R. and Clark J.W. Solid St. Comm.
{\bf 96}, 353 (1995).
\item [17)] Abrikosov A.A., Campuzano J.C. and Gofron K. Physica {\bf
C214}, 73 (1993).
\end{itemize}
\newpage
\begin{center}
{\bf Figure Captions}
\end{center}
\vskip 0.3truecm
\noindent

Fig.1. Schematic view of a quasiparticle dispersion with a downswing at
the Fermi energy and the corresponding Maxwell construction.
\medskip
\noindent

Fig.2. Normal $(n_p)$ and anomalous $(\kappa_p)$ occupation numbers
in the model with the Fermion condensate.
\medskip
\noindent

Fig.3.  The quasiparticle energy $E_p$ (eq.(30)) as
a function of temperature  (in units of $V_1$).
\medskip
\noindent

Fig.4. The gap function $\Delta_p=\sqrt{E^2_p-\xi^2_p}$ as a function
of $\xi_p=\varepsilon_p-\mu$ for various temperatures.
\medskip
\noindent

Fig.5. The quasiparticle dispersion around the chemical potential
for various temperatures $T\geq T_c$.
\medskip
\noindent

Fig.6. The level density
$\rho(\varepsilon)$ corresponding to the same situation as
in Fig.5.
\end{document}